\documentclass{PoS}

\usepackage{cite}

\title{Spatial structure of the color field in the SU(3) flux tube}

\ShortTitle{Spatial structure of the color field in the SU(3) flux tube}

\author{M. Baker \\
Department of Physics, University of Washington, WA 98105 Seattle,
USA \\
E-mail: \email{mbaker4@uw.edu}}

\author{Paolo Cea \\
INFN - Sezione di Bari, I-70126 Bari, Italy \\
E-mail: \email{paolo.cea@ba.infn.it}}

\author{\speaker{Volodymyr Chelnokov} \\
INFN - Gruppo collegato di Cosenza, I-87036 Arcavacata di Rende,
  Cosenza, Italy, \\
on leave of absence from Bogolyubov Institute for Theoretical Physics of
National Academy of Sciences of Ukraine\\
E-mail: \email{volodymyr.chelnokov@lnf.infn.it}}

\author{Leonardo Cosmai \\
INFN - Sezione di Bari, I-70126 Bari, Italy \\
E-mail: \email{leonardo.cosmai@ba.infn.it}}

\author{Francesca Cuteri \\
Institut f\"ur Theoretische Physik, Goethe Universit\"at,
            60438 Frankfurt am Main, Germany\\
E-mail: \email{cuteri@th.physik.uni-frankfurt.de}}

\author{Alessandro Papa \\
Dipartimento di Fisica, Universit\`a della Calabria,
I-87036 Arcavacata di Rende, Cosenza, Italy \\
and INFN - Gruppo collegato di Cosenza, I-87036 Arcavacata di Rende, Cosenza,
Italy \\
E-mail: \email{papa@cs.infn.it}}

\abstract{We report on the chromoelectric and chromomagnetic fields generated
by a static quark-antiquark pair at zero temperature in pure gauge SU(3).
From the spatial structure of chromoelectric field we extract
its nonperturbative part and discuss its properties.}

\FullConference{The 36th Annual International Symposium on Lattice Field Theory - LATTICE2018\\
		22-28 July, 2018\\
		Michigan State University, East Lansing, Michigan, USA.}

\begin{document}

\section{Introduction}

Monte Carlo simulations of the lattice theories provide a way to investigate
the nonperturbative behavior of strongly interacting particles.
Numerical studies of various effective
models~\cite{Fukugita:1983du,Kiskis:1984ru,Flower:1985gs,Wosiek:1987kx,DiGiacomo:1989yp,DiGiacomo:1990hc,Cea:1992sd,Matsubara:1993nq,Cea:1994ed,Cea:1995zt,Bali:1994de,Skala:1996ar,Haymaker:2005py,D'Alessandro:2006ug,Cardaci:2010tb,Cea:2012qw,Cea:2013oba,Cea:2014uja,Cea:2014hma,Cardoso:2013lla,Caselle:2014eka,Cea:2017ocq,Shuryak:2018ytg}
show that quark confinement is realized by the formation of ``flux tubes'',
in which the chromoelectric field created by a quark-antiquark pair is confined
to a narrow region around the line connecting the quarks,
which provides linear potential between the quarks~\cite{Bander:1980mu,Greensite:2003bk,Ripka:2005cr,Simonov:2018cbk}. The field in such flux tubes is
mainly directed along the tube axis and does not change when moving along the axis far enough from the sources~\cite{Cea:1995zt}.
Any theory claiming to explain the mechanism of confinement should be able to describe the
field distribution in the flux tube in good agreement with the lattice data.

We report on measurements of all the components of the chromoelectromagnetic field created by a
static quark-antiquark pair in pure gauge SU(3) at zero temperature, for different values of quark-antiquark separation.
Our results show that while the longitudinal field component is dominant,
the behavior of the transverse components can be used to subtract the perturbative part
of the field, getting
what we identify as the confining field forming the  flux tube.

\section{Lattice observables and details of measurements}

We performed our simulations for a pure SU(3) lattice gauge theory in four
dimensions, with standard Wilson action. The simulations were performed using the
publicly available MILC code, modified to introduce the relevant observables.

\begin{figure}[htb]
\centering
\includegraphics[scale=0.4,clip]{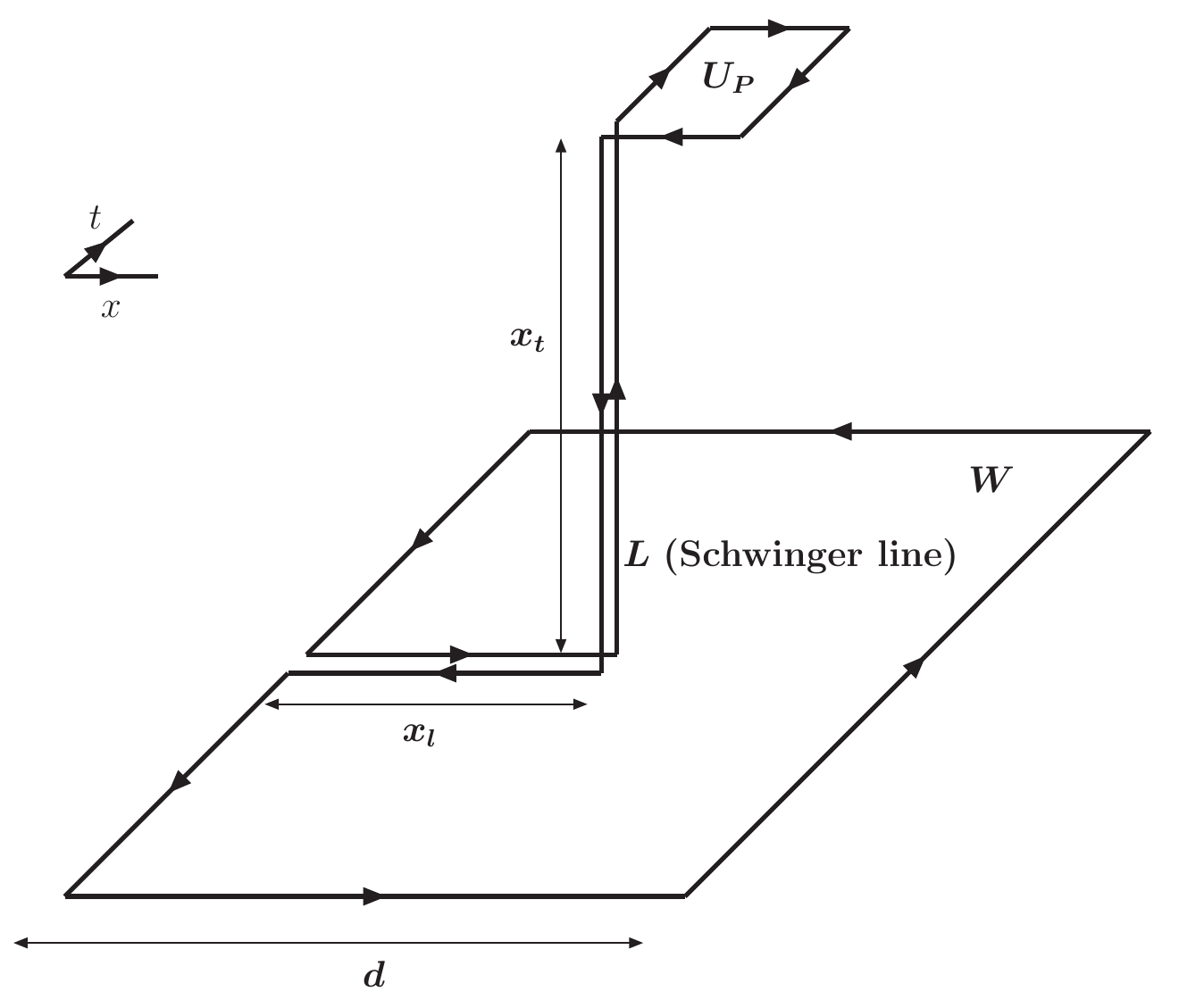}
\caption{The connected correlator given in Eq.~(\protect\ref{rhoW})
between the plaquette $U_{P}$ and the Wilson loop
(subtraction in $\rho_{W,\,\mu\nu}^{\rm conn}$ not explicitly drawn).}
\label{fig:op_W}
\end{figure}

To measure the field generated by a quark-antiquark pair we use
the connected correlation function~\cite{DiGiacomo:1989yp,DiGiacomo:1990hc,Kuzmenko:2000bq,DiGiacomo:2000va}

\begin{equation}
\label{rhoW}
\rho_{W,\,\mu\nu}^{\rm conn} = \frac{\left\langle {\rm tr}
\left( W L U_P L^{\dagger} \right)  \right\rangle}
              { \left\langle {\rm tr} (W) \right\rangle }
 - \frac{1}{3} \,
\frac{\left\langle {\rm tr} (U_P) {\rm tr} (W)  \right\rangle}
              { \left\langle {\rm tr} (W) \right\rangle } \; ,
\end{equation}
where $U_P$ is a plaquette in the $\mu \nu$ plane, $W$ is a Wilson loop that
generates a static quark-antiquark pair, and $L$ is a Schwinger line connecting
the plaquette to the Wilson loop (see Figure~\ref{fig:op_W}). The connected
correlation function (\ref{rhoW}) has a naive continuum limit~\cite{DiGiacomo:1990hc}
\begin{equation}
\label{rhoWlimcont}
\rho_{W,\,\mu\nu}^{\rm conn}\stackrel{a \rightarrow 0}{\longrightarrow} a^2 g
\left[ \left\langle
F_{\mu\nu}\right\rangle_{q\bar{q}} - \left\langle F_{\mu\nu}
\right\rangle_0 \right]  \;,
\end{equation}
where the subscript $q\bar{q}$ refers to the average field in presence of a
quark-antiquark pair and the subscript $0$ denotes average value in vacuum, which is
expected to vanish. Different orientations of the plaquette correspond to the
three components of the chromoelectric field $\vec{E}$, when
the plaquette is time-like, and to the three components of
the chromomagnetic field $\vec{B}$, when the plaquette is spatial.

To reduce fluctuations we used a smearing procedure consisting of one HYP smearing
step~\cite{Hasenfratz:2001hp}
with parameters $(\alpha_1, \alpha_2, \alpha_3) = (1.0, 0.5, 0.5)$
for the links in the temporal direction, and a set of $N_{\mathrm{sm}}$
APE smearing steps~\cite{Falcioni1985624} for the links in spatial directions with
$\alpha_{\mathrm{APE}} = 0.25$.

We have performed simulations for the three different values of
$\beta$, described in Table~{\ref{tbl:runs}}.
To set the physical scale a parametrization given in ~\cite{Edwards:1998xf} was used,
using the value $\sqrt{\sigma} = 420 ~$MeV for the string tension.
The measurements in each case
were taken every 100 lattice updates, discarding a few thousands
trajectories at the beginning to reach thermalization. Error analysis was made
using jackknife algorithm, with varying blocking size.

\begin{table}[htb]
\centering
\begin{tabular}{|c|c|c|c|c|}
\hline\hline
$\beta$ & lattice & $d$ [fm] & statistics & smearing steps \\ \hline
6.370  &         & 0.951(11)  &  5300  & 100  \\
6.240  & $48^4$  & 1.142(15)  &  21000 & 100  \\
6.136  & 	 & 1.332(20)  &  84000 & 120  \\
\hline\hline
\end{tabular}
\caption{Summary of the simulation parameters.}
\label{tbl:runs}
\end{table}

\section{Measurement results and Coulomb field subtraction}

At each value of $\beta$ we measured all six components of the fields at the set of
displacements $(x_l, x_t)$ from the static quark (see Figure~\ref{fig:op_W}).
The equivalence of the two choices of the direction of $x_t$ perpendicular to $x_l$,
coming from the rotational symmetry of the system, was explicitly checked numerically.
In what follows, the $x$ axis is chosen along the quark-antiquark line,
and the $y$ axis along the direction of transverse displacement.

\begin{figure}[htb]
\centering
\includegraphics[width=0.495\textwidth,clip]{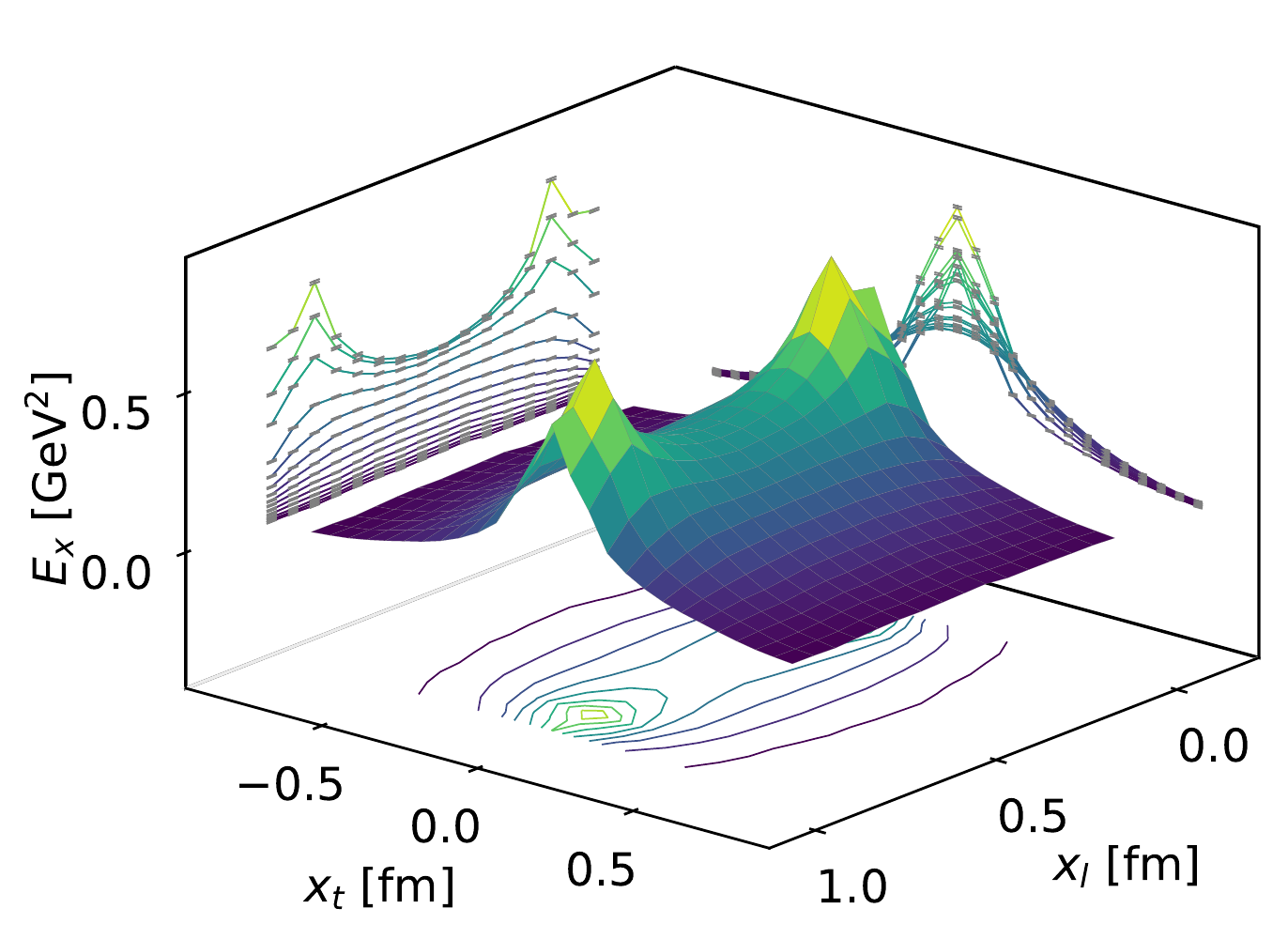}
\includegraphics[width=0.495\textwidth,clip]{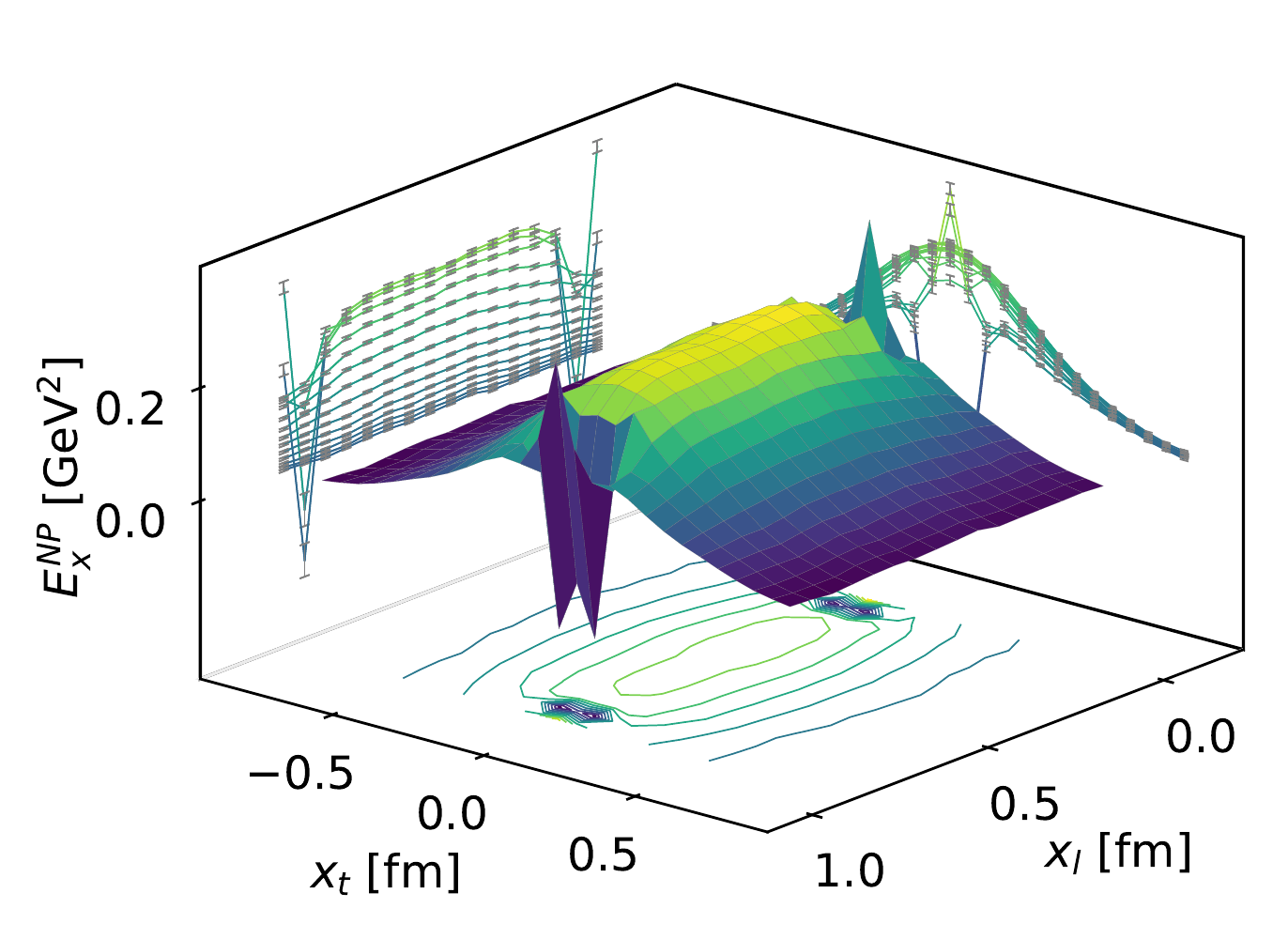} \\
\includegraphics[width=0.495\textwidth,clip]{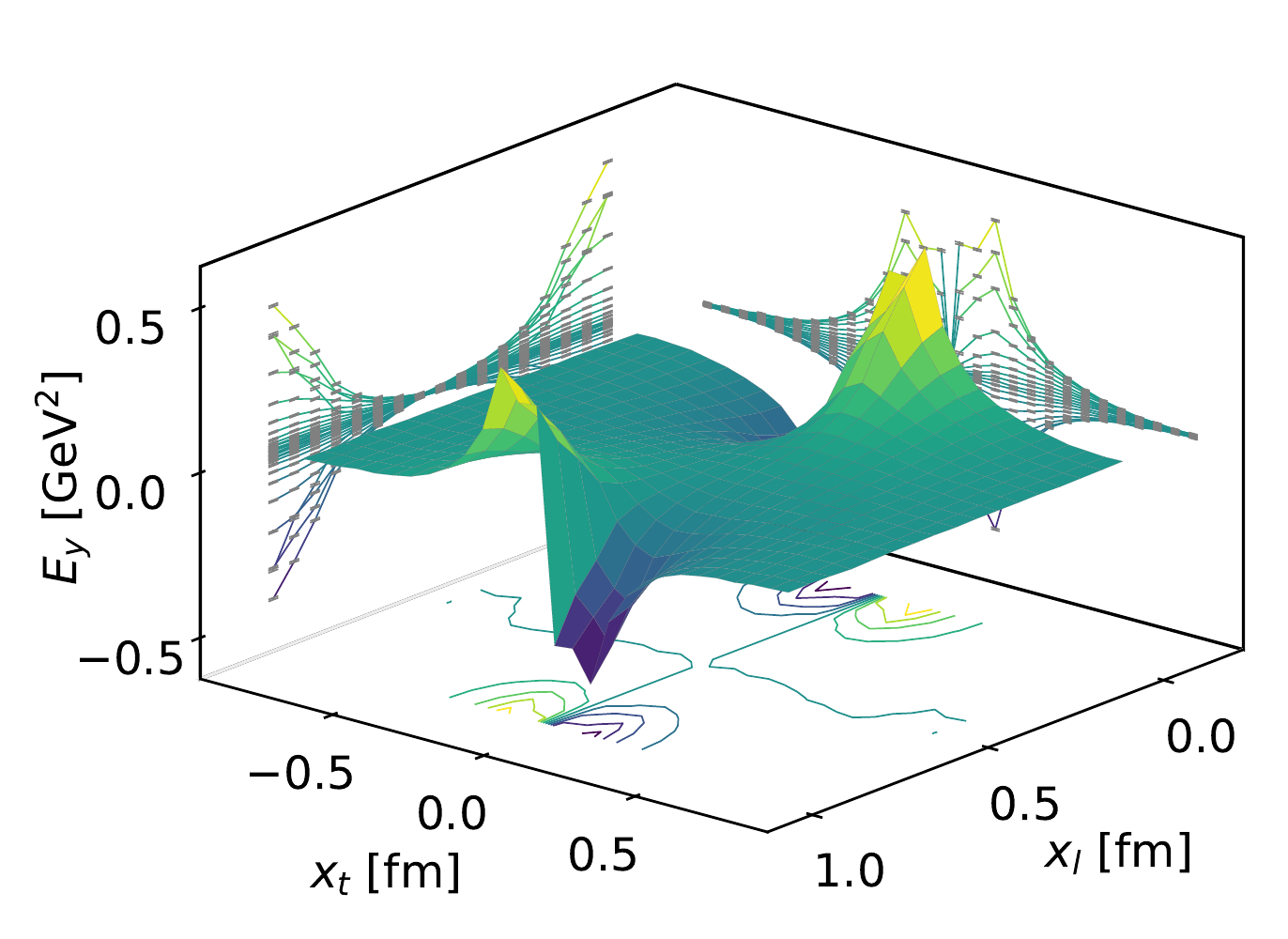}
\includegraphics[width=0.495\textwidth,clip]{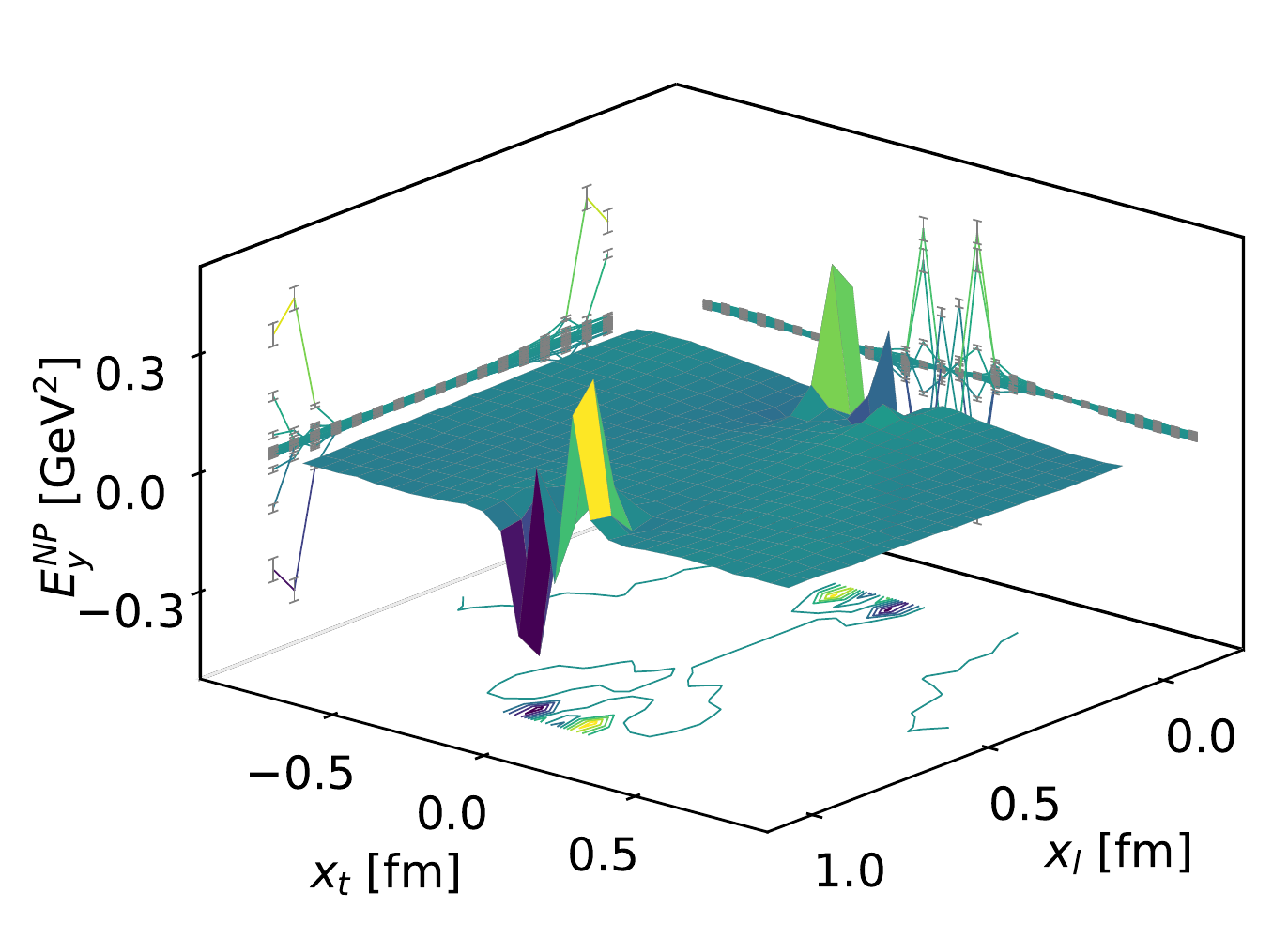} \\
\includegraphics[width=0.495\textwidth,clip]{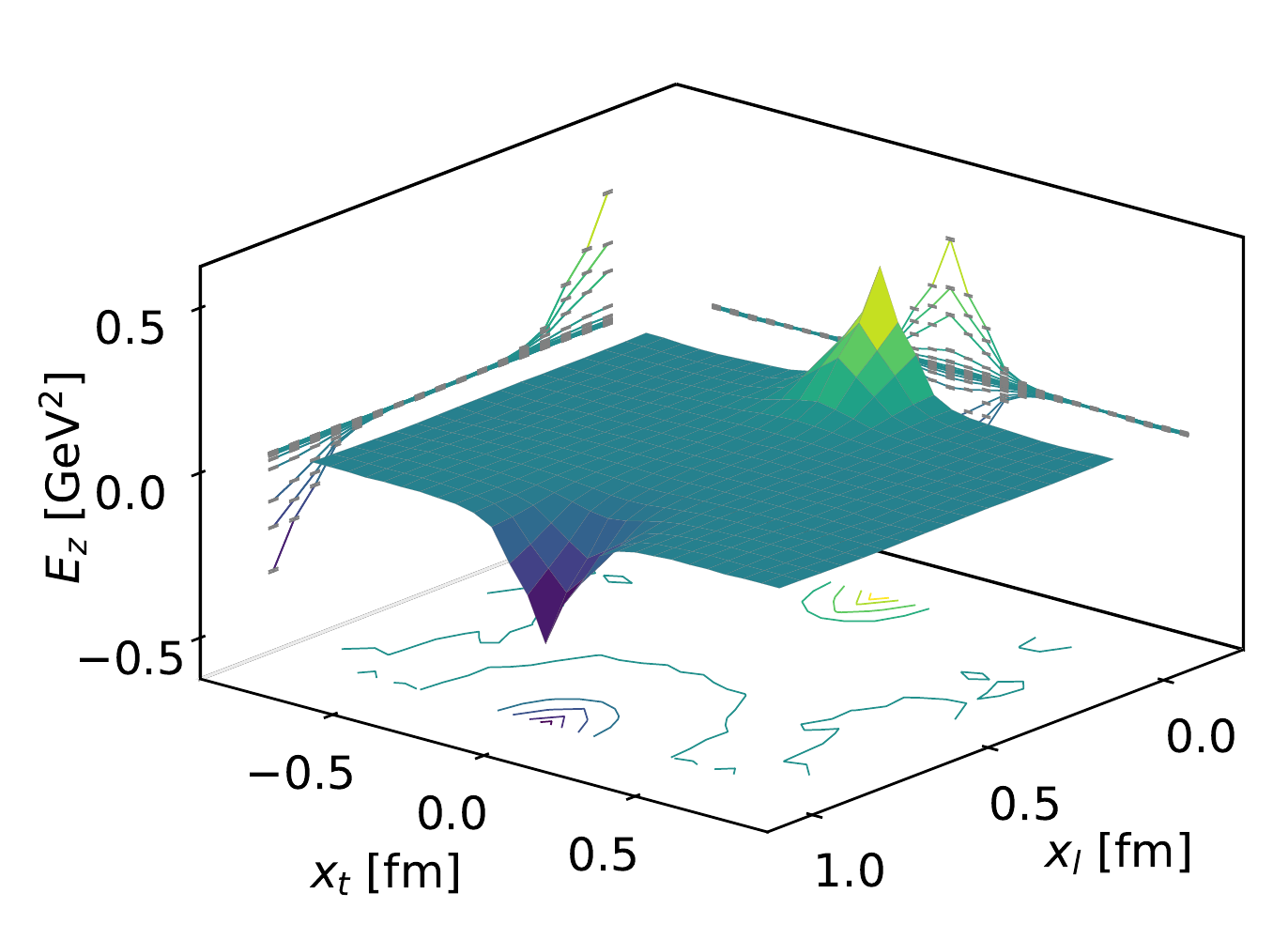}
\includegraphics[width=0.495\textwidth,clip]{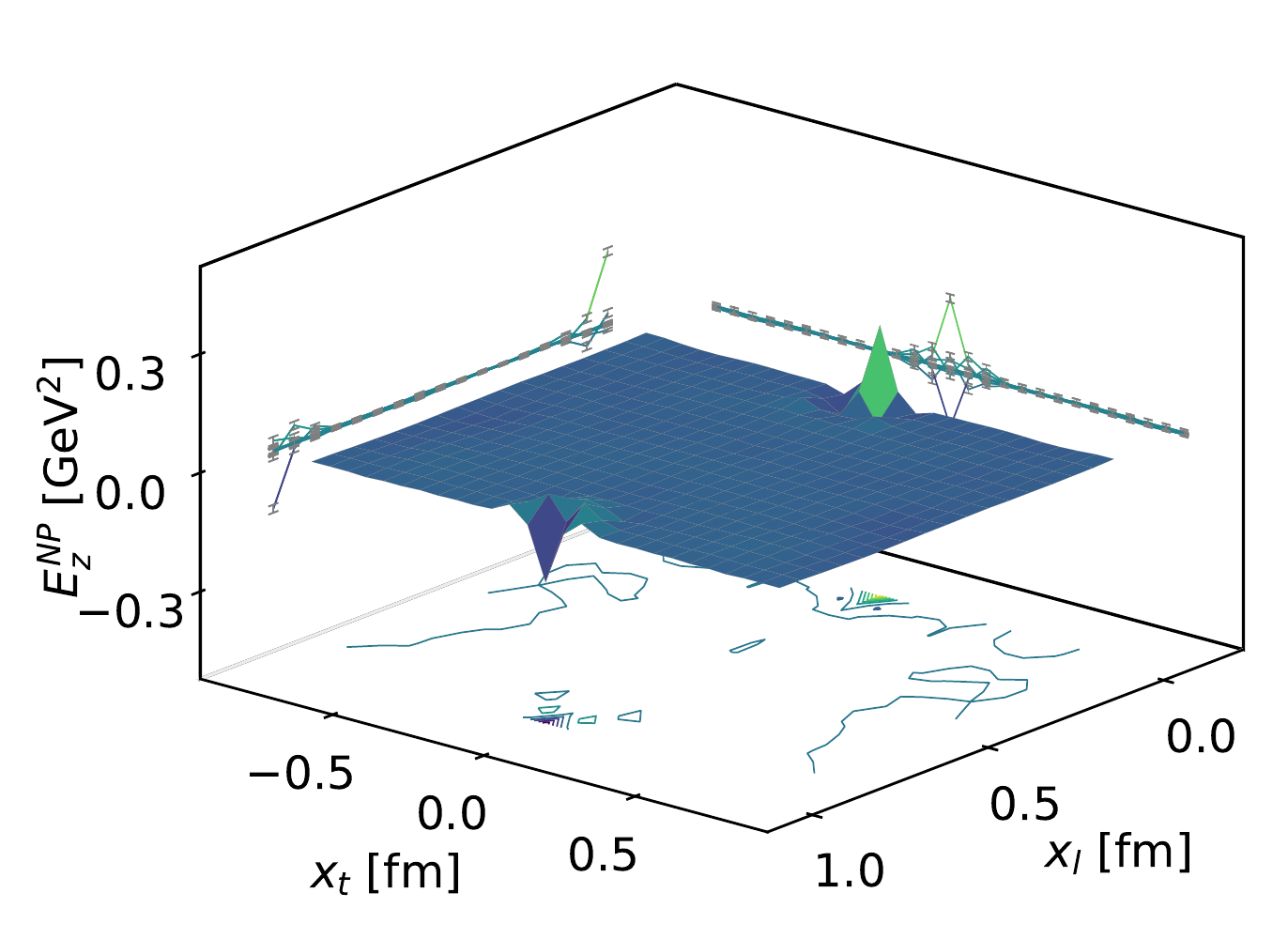}
\caption{Surface and contour plots for the three components of the
  chromoelectric field (left) and its nonperturbative part (right) at $\beta=6.370$ and $d=0.951$ fm.}
\label{fig:Fields}
\end{figure}

The three components of the chromomagnetic field $\vec{B}$ are compatible
with zero within errors. The values of the three components of the chromoelectric field $\vec{E}$ at
$\beta = 6.370$ as a function of the displacements $x_l$ and $x_t$ are shown on the left in Figure~\ref{fig:Fields}.
The transverse components of the chromoelectric field
can be nicely described as a Coulomb-like field $\vec{E}^{C}$
\begin{equation}
\label{C2}
 \vec{E}^C (\vec{r}; Q, R_0) \; = \;  Q \left( \frac{\vec{r}_1}{\max(r_1, R_0)^3} \; - \;  \frac{\vec{r}_2}{\max(r_2, R_0)^3} \; \right ) \; ,
\end{equation}
where $Q$ is an effective charge, $\vec{r}_1$ and $\vec{r}_2$ are vectors from the positions of the static quark
and antiquark, respectively, to the point $\vec{r}$ where the field is measured.
Since the field components are probed by a plaquette, we consider $\vec{r}$ to be the radius vector to the center of this plaquette.
One important consequence of this is that the field $E_z$ is measured at points that are
a half lattice spacing away from the $x y$ plane, so $E_z^C$ is not equal to zero.
Finally $R_0$ is an effective radius of the charge, which we introduced to
try to describe the behavior of the field on distances less then three lattice spacing from the quarks --
where the field $\vec{E}^C$ fails to give a good description of the measured field.

We performed a series of fit of the two transverse field components $E_y$ and $E_z$
at different planes transverse to the line connecting
the two quarks, treating the effective charge $Q$ and effective radius $R_0$ as the fit parameters.
The results of these fits are summarized in Table~\ref{tbl:coulombFit}; the
errors of fit parameters were obtained from a comparison of the best fit parameters
for different transverse planes at which we managed to obtain a good fit.
\begin{table}[htb]
\centering
\begin{tabular}{|c|c|c|c|c|}
\hline\hline
$\beta$ & $Q$ & $R_0/a$ & $R_0$ [fm] & $d$ [fm] \\ \hline
6.370  & 0.278(4)  &  1.920(14) &  0.1142(16) & 0.951(11) \\
6.240  & 0.289(11) &  1.92(3)   &  0.1367(29) & 1.142(15) \\
6.136  & 0.305(14) &  2.15(7)   &  0.179(6)   & 1.332(20) \\
\hline\hline
\end{tabular}.
\caption{Values of the fit parameters extracted from Coulomb fits of the transverse
components of the chromoelectric field.}
\label{tbl:coulombFit}
\end{table}

\begin{figure}[htb]
\centering
\includegraphics[width=0.495\textwidth,clip]{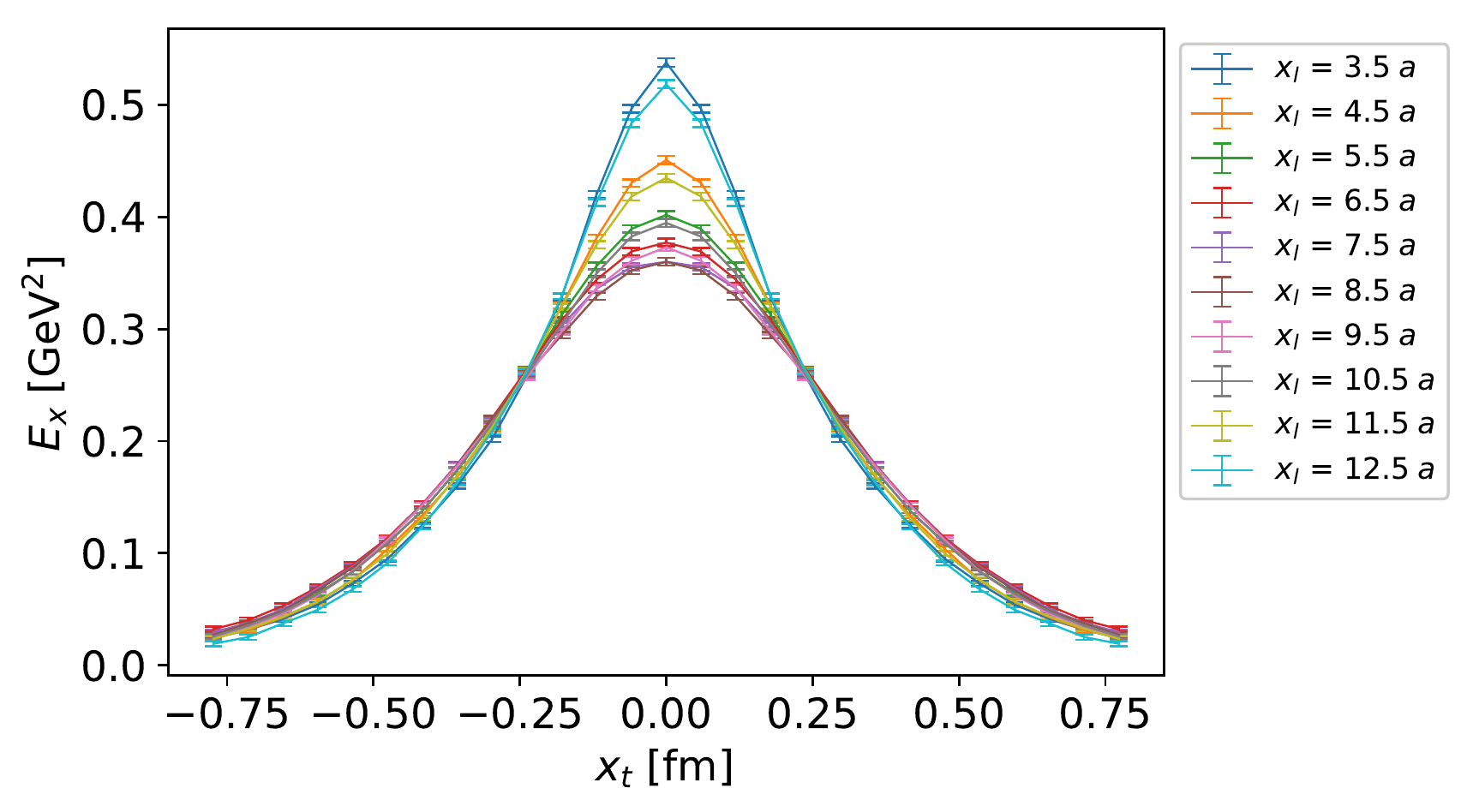}
\includegraphics[width=0.495\textwidth,clip]{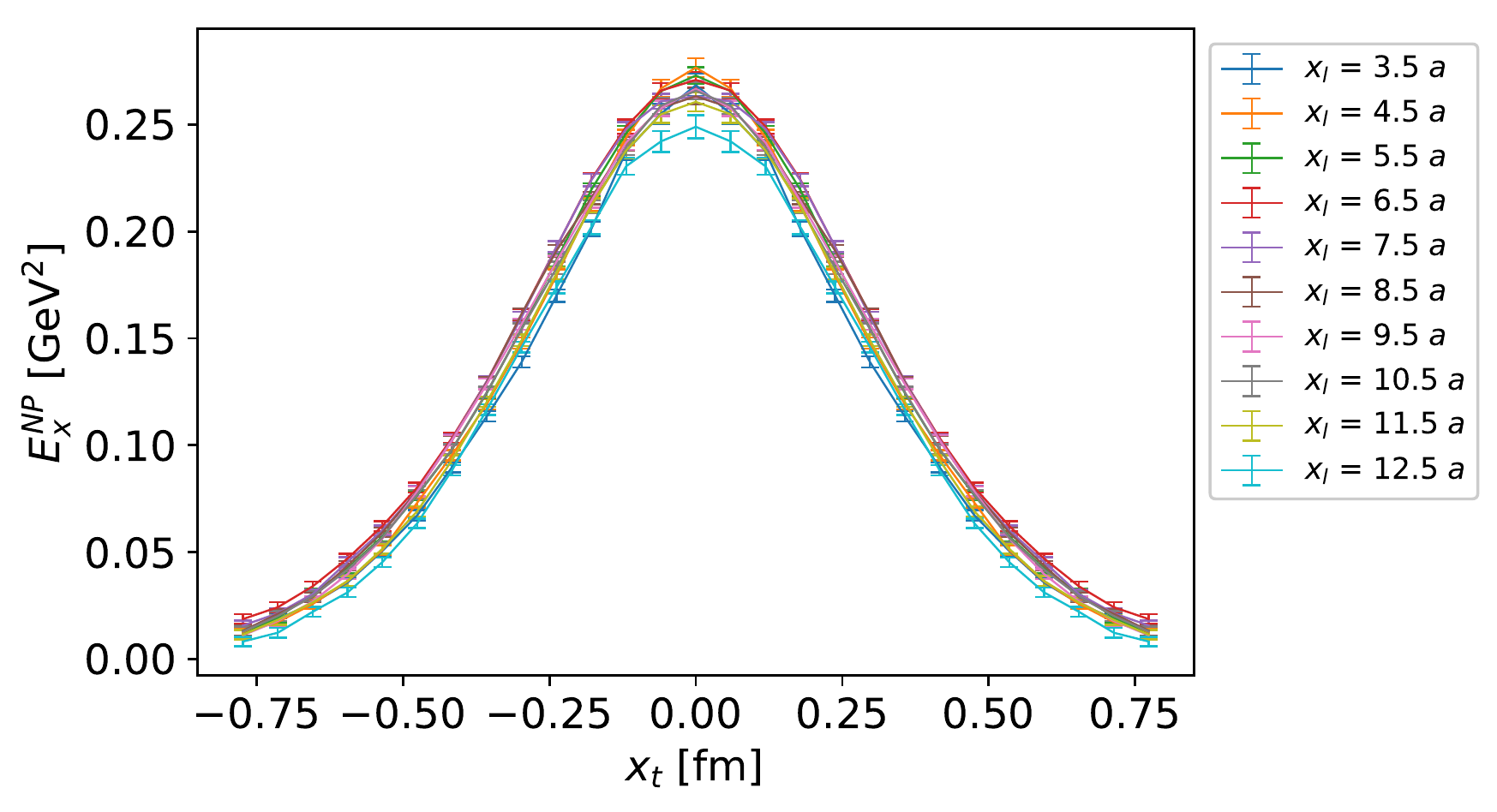} \\
\includegraphics[width=0.495\textwidth,clip]{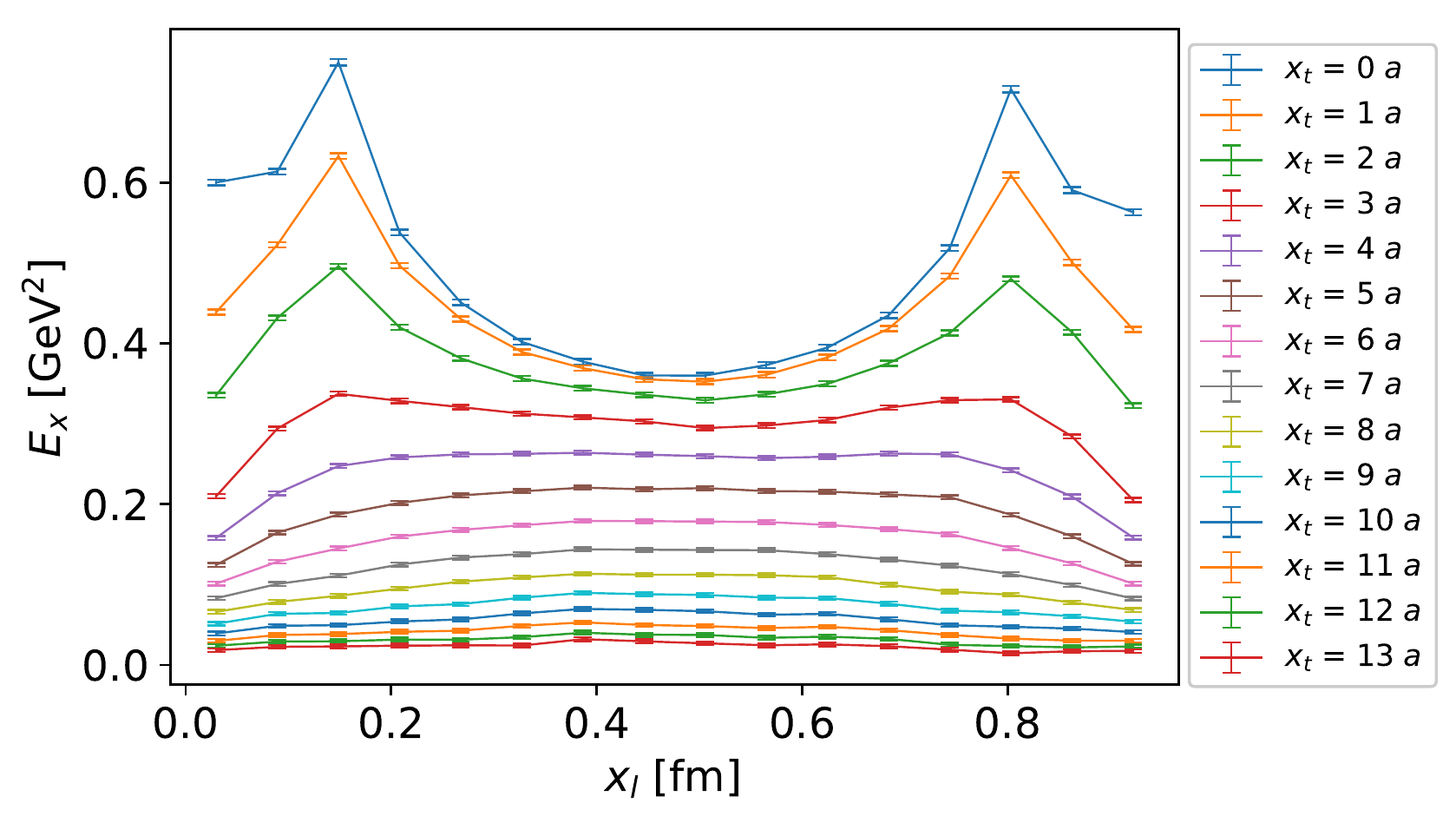}
\includegraphics[width=0.495\textwidth,clip]{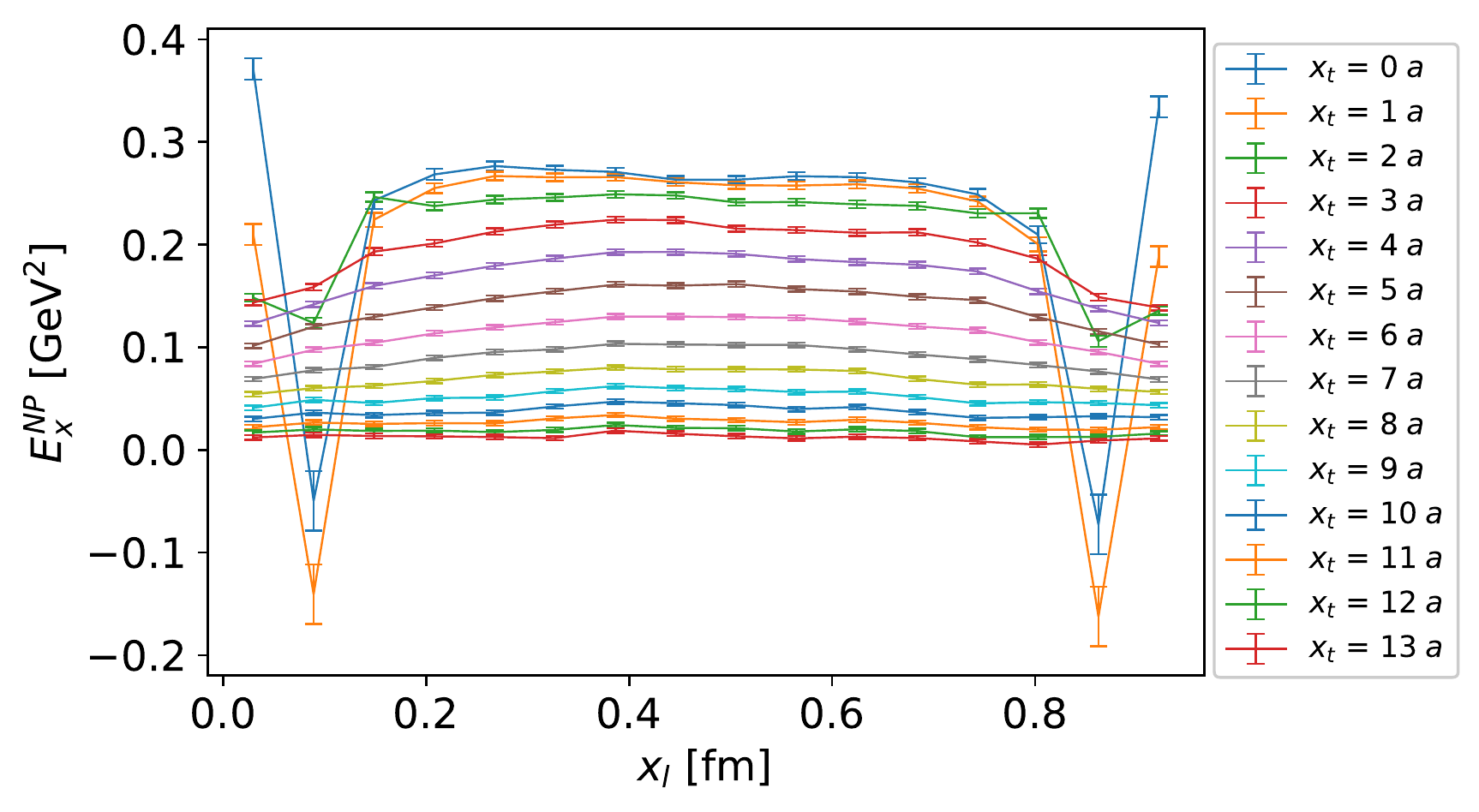}
\caption{Transverse (top) and longitudinal (bottom) sections of the longitudinal component of the full
chromoelectric field (left) and its nonperturbative part (right) at $\beta=6.370$ and $d=0.951$ fm. }
\label{fig:sections}
\end{figure}

In the further analysis we consider the chromoelectric field $\vec{E}$ to consist of
two parts - nonperturbative part $\vec{E}^{NP}$ and the perturbative Coulomb-like part $\vec{E}^C$,
\begin{equation}
 \vec{E}(\vec{r}) = \vec{E}^{NP}(\vec{r}) + \vec{E}^{C}(\vec{r}; Q, R_0) \; ,
\end{equation}
taking the values of $Q$ and $R_0$ from the results of our previous fitting analysis.
In this way we extract the nonperturbative contribution to the chromoelectric field.
This procedure makes the nonperturbative part purely longitudinal,
apart from the small region of radius ${}\sim R_0$ around the static quarks.
The three components of the obtained nonperturbative field $\vec{E}^{NP}$ for $\beta = 6.370$ are shown
on the right in Figure~\ref{fig:Fields}.
As can be seen from these plots the
longitudinal component of the field after subtraction $E_x^{NP}$ is much more stable
under changes of the $x_l$ coordinate, making it useful to extract the shape of the flux tube
for small distances between quarks.
This fact is more clear when looking at transverse and longitudinal sections of the field
shown in Figure~\ref{fig:sections}.
We remark that while all the plots are given for the
$\beta = 6.370$ case the results are qualitatively the same for other $\beta$ values.

\section{Conclusions}

We have performed measurements of the six components of the chromoelectromagnetic
field created by a static quark-antiquark pair in the four-dimensional pure gauge SU(3) theory at zero temperature.

The measured chromomagnetic field is compatible with zero.
The transverse components of the chromoelectric field decay fast with the distance from the two quarks,
and can be described by the Coulomb-like behavior. Extracting the parameters of the Coulomb law
from the transverse field components, and subtracting  the longitudinal component of the resulting Coulomb-like field
from the measured longitudinal chromoelectric field, yields, in a model-independent way,
the nonperturbative chromoelectric field forming the confining flux tube, thereby removing
(to a certain degree)
the perturbative short-range corrections, existing in the original field.
We believe this approach can be straightforwardly applied also at nonzero temperature and/or
within a theory with dynamical quarks.

\section{Acknowledgements}
This investigation was in part based on the MILC collaboration's public
lattice gauge theory code. See
{\href{http://physics.utah.edu/~detar/milc.html}{http://physics.utah.edu/~detar/milc.html}}.
Numerical calculations have been made possible through a CINECA-INFN
agreement, providing access to resources on MARCONI at CINECA.
AP, LC, PC, VC acknowledge support from INFN/NPQCD project.
FC acknowledges support from the German Bundesministerium f{\"u}r Bildung und Forschung (BMBF)
under Contract No. 05P1RFCA1/05P2015 and from the DFG (Emmy Noether Programme EN 1064/2-1).
VC acknowledges financial support from the INFN HPC{\_}HTC project.

\bibliographystyle{JHEP}

\providecommand{\href}[2]{#2}\begingroup\raggedright\endgroup

\end{document}